\documentclass[11pt,a4paper]{article}\pdfoutput=1
\usepackage{amsmath,amsfonts,amssymb,amscd,float}
\usepackage{cite}
\usepackage[small,bf,hang]{caption}
\usepackage{setspace}
\usepackage{tensor}
\usepackage{slashed}
\usepackage{amsmath}
\usepackage{latexsym,epsfig}
\usepackage{arydshln}
\usepackage{extarrows}
\usepackage[utf8]{inputenc}
\usepackage[linktoc=all,colorlinks,citecolor=blue]{hyperref}
\usepackage{amsthm,tikz-cd,accents}
\usepackage{bm}
\usepackage{multicol}
\usepackage{scalerel}
\usepackage{stackengine,wasysym}
\usepackage{tcolorbox}

\newcommand\lf{$L_\infty${}}

\newcommand\maps{{\rm Maps}}
\newcommand{\into}{\hookrightarrow}
\newcommand{\onto}{\twoheadrightarrow}

\newcommand{\bbR}{\mathbb{R}}

\newcommand{\dr}{\mathrm{d}}

\usepackage[makeroom]{cancel}

\usepackage{color}

\def\alex{
\topmargin -60pt
\oddsidemargin +5pt
\headheight 5pt \headsep 0pt
\textwidth 6.0in 
\textheight 10.85in 
\marginparwidth .875in
\parskip 5pt plus 1pt \jot = 1.5ex}
\usepackage{lmodern}
\usepackage[T1]{fontenc}

\usepackage{microtype}
\alex
 \csname
@addtoreset\endcsname{equation}{section}

\def\moth{\mathsurround=0pt}
\newdimen\zo \zo=0pt

\def\tick{\leaders\hrule height 0.5ex depth 0pt \hskip 0.5pt}
\def\upboxfill{$\moth \setbox\zo\hbox{\tick}%
\hskip 3pt\hbox to 0pt{$\tick$\hss}\hrulefill \hbox to 7.5pt{$\tick$\hss}$}

\def\dtick{\leaders\hrule height .34pt depth 0.5ex \hskip 0.5pt}
\def\downboxfill{$\moth \setbox\zo\hbox{\dtick}%
\hskip 2pt\hbox to 0pt{$\dtick$\hss}\hrulefill \hbox to 2pt{$\dtick$\hss}$}

\def\pd{\partial}

\def\bec{\begin{center}}
\def\ec{\end{center}}

\def\S{\Sigma}

\def\cM{{\cal M}}

\def\del{\partial}

\def\be{\begin{equation}}
\def\ee{\end{equation}}
\def\bea{\begin{eqnarray}}
\def\eea{\end{eqnarray}}
\def\ba{\begin{array}}
\def\ea{\end{array}}

\definecolor{cambridgeblue}{rgb}{0.64, 0.76, 0.68}
\definecolor{lapislazuli}{rgb}{0.15, 0.38, 0.61}
\definecolor{awesome}{rgb}{1.0, 0.13, 0.32}
\definecolor{aureolin}{rgb}{0.99, 0.93, 0.0}
\definecolor{almond}{rgb}{0.94, 0.87, 0.8}
\definecolor{antiquewhite}{rgb}{0.98, 0.92, 0.84}

\usepackage[left, final]{showlabels}
\begin{document}

\begin{titlepage}
	\rightline{\today}\vskip0.5em
	\rightline{HU-EP-22/02}
	\begin{center}
		{\large \bf{Romans massive QP manifolds}
		}\\
		\vskip 1.0cm

		{\large {Alex S.~Arvanitakis$^a$, Emanuel Malek$^b$, David Tennyson$^c$}}
		\vskip .3cm
		
		{\it  
			$^a$Theoretische Natuurkunde, Vrije Universiteit Brussel, \\ and the International Solvay Institutes, \\ Pleinlaan 2, B-1050 Brussels, Belgium}
		
		{\it
		    $^b$ Institut f\"ur Physik, Humboldt-Universit\"at zu Berlin, IRIS Geb\"aude, \\ Zum Gro{\ss}en Windkanal 2, 12489 Berlin, Germany}
		    
		{\it
		    $^c$Department of Physics, Imperial College London, \\ Prince Consort Road, London, SW7 2AZ, UK}

		\vskip .3cm
		\texttt{alex.s.arvanitakis@vub.be}\\
		\texttt{emanuel.malek@physik.hu-berlin.de}\\
		\texttt{d.tennyson16@imperial.ac.uk}

		\vskip 0.3cm
		{\bf Abstract}
		
	\end{center}
	\begin{narrower}
	\noindent
	We introduce QP manifolds that capture the generalised geometry of type IIA string backgrounds with Ramond-Ramond fluxes and Romans mass.	Each of these is associated to a BPS brane in type IIA: a D2, D4, or NS5-brane. We explain how these probe branes are related to their associated QP-manifolds via the AKSZ topological field theory construction and the recent brane phase space construction. M-theory/type IIA duality is realised on the QP-manifold side as symplectic reduction along the M-theory circle (for branes that do not wrap it);  this always produces IIA QP-manifolds with vanishing Romans mass.
	\end{narrower}

\end{titlepage}
	\newpage
	{\hypersetup{linkcolor=black}\tableofcontents}

\section{Introduction}

It is well established that the natural language to describe backgrounds of string theory and M-theory is not through conventional geometry, but through higher `algebroids' \cite{baraglia2012leibniz}.The NS sector, for example, can be described by a Courant algebroid on a bundle $T\oplus T^{*}$ \cite{Coimbra:2011nw}. If one includes the RR sector as well then one must work on a slightly more general Leibniz algebroid \cite{Hull:2007zu,PiresPacheco:2008qik,Coimbra:2012af,Cassani:2016ncu}. These higher algebroids, called generalised geometries in the physics community, unify both the enhanced global symmetry of the background along with the enhanced gauge symmetries that come from combining diffeomorphisms and $n$-form gauge transformations. They are characterised by a `generalised tangent bundle' $E\rightarrow M$, sections  $V$ of which  generate the infinitesimal gauge transformations through a derivation called the Dorfman or generalised Lie derivative $L_{V}$. This derivative satisfies a Leibniz identity but is not antisymmetric, unlike the Lie derivative. Instead, it is only antisymmetric up to homotopy - a key property of these higher algebroids. Given a generalised tangent bundle, the choice of Dorfman derivative is not unique but can be twisted by a set of fluxes satisfying the Bianchi identities. 

It has been noted that, in many cases, these algebroids have an equivalent description in terms of a particular class of non-commutative geometries called QP manifolds (see e.g. \cite{Cattaneo:2010re} for a review). These are graded supermanifolds\footnote{A supermanifold with an extra $\mathbb Z$-grading on its structure sheaf, so all functions $f$ have a well-defined degree $\deg f$. For us $\deg f\geq 0$ and $\deg\mathrm{mod}\; 2$ equals the Grassmann parity. This grading is equivalent to the existence of a degree-counting vector field (``Euler'') which we employ later.} $\mathcal{M}$ that come equipped with a degree $-P$ Poisson bracket $(\cdot,\cdot)$ and a degree 1 vector field $Q$ such that $Q^{2} = 0$. It turns out that one can write the vector field $Q=(\Theta,\cdot)$ for some homogeneous Hamiltonian function $\Theta$ of degree $P+1$. The condition $Q^{2} = 0$ is then equivalent to $(\Theta,\Theta)=0$, the classical master equation.

The link between QP manifolds and higher algebroids was first noted in \cite{roytenberg1999courant} where they showed that any Courant algebroid $E\rightarrow M$ could be described in terms of the degree 2 graded symplectic manifold $\mathcal{M} = T^{*}[2]T[1]M$. Both the anchor map\footnote{The anchor map $a:E\rightarrow T$ is intrinsic to the definition of these higher algebroids. See e.g. \cite{baraglia2012leibniz}.} and the possible flux twistings are encoded in the choice of $\Theta$, sections of $E$ are given by homogeneous functions of degree 1, and the generalised Lie derivative is given by $((\Theta,V),V')$. In \cite{Arvanitakis:2018cyo}, this correspondence was extended to include the Leibniz algebroids relevant for backgrounds of M-theory and type IIB by considering higher degree graded symplectic manifolds. There, it was also noted that homogeneous functions on $\mathcal{M}$ correspond to sections of the bundles appearing in the tensor hierarchy \cite{deWit:2008ta} of exceptional field theory, and that $Q$ is the QP equivalent of the derivative $\hat{\del}$. In this paper we complete this correspondence by finding the QP manifolds relevant for IIA algebroids, both massive and massless.

Not only does the QP picture provide a unified description of the algebroids and the tensor hierarchies of string/M-theory, but it also helps to uncover the relationship between these algebroids and brane physics. For example, the AKSZ construction \cite{Alexandrov:1995kv} allows one to construct a topological theory on a manifold of dimension $P+1$ from a degree $P$ QP manifold. When performed on the space $\mathcal{M}=T^{*}[2]T[1]M$, one finds the Wess-Zumino (WZ) term for the string, i.e. the integral over a 3-manifold $\Sigma_{3}$ of the NS 3-form flux $H$, where the boundary $\del\Sigma_{3}$ is interpreted as the worldvolume of the string. In \cite{Arvanitakis:2018cyo} this was extended to the M2, M5 and D3 branes, where it was shown that the AKSZ construction reproduced the relevant WZ term. The QP structure also allows one to study the phase space and current algebras for branes which was done in \cite{Ikeda:2011ax,Alekseev:2004np} for the fundamental string and M2 brane and in \cite{Arvanitakis:2021wkt} for M5 and D3 branes. In this paper we will outline how one can extend this analysis further to include the D2, D4 and NS5 branes of type IIA, including a Romans mass \cite{Romans:1985tz} deformation for the D2 and D4; we treat the D2 brane in detail. 

This paper is structured as follows. In section \ref{sec:QP_for_IIA}, we write down the QP manifolds associated to the D2, D4 and NS5 branes and derive the most general Hamiltonian function $\Theta$ in each case. We find that the D2 and D4 QP manifolds allow for a Romans mass, while the NS5 does not. We also show that the condition $(\Theta,\Theta)=0$ precisely recovers the Bianchi identities, and hence $\Theta$ encodes all of the possible flux-twistings of the associated Leibniz algebroid. We further show that homogeneous functions of degree $P-1$ can be interpreted as sections of the generalised tangent bundle for (massive) IIA, and the bracket defined by $((\Theta,V),V')$ precisely recovers the flux-twisted generalised Lie derivative of \cite{Cassani:2016ncu}. In section \ref{sec:M/IIA_duality} we show how one can recover the NS5 and D2 QP manifolds from a circle reduction of the QP manifolds of M5 and M2 branes respectively. In section \ref{sec:applications}, we derive the AKSZ topological model for the QP manifolds and show that they recover the WZ term for the respective branes, hence justifying the nomenclature. We also study the brane phase space for the D2 brane. Finally, section \ref{sec:discussion} is left for discussion. For conventions and useful formulae, see e.g. \cite[Appendix A, B]{Arvanitakis:2018cyo}.

\section{QP manifolds for IIA branes}\label{sec:QP_for_IIA}

\subsection{D2 [$P=3$ algebroid]}

The QP manifold for the D2 brane is the following
\begin{equation}
	\mathcal{M} = T^{*}[3]T[1]M \times \bbR[1]\times \bbR[2] \,.
\end{equation}
It has the following graded coordinates with a natural degree 3 symplectic form $\omega$ which induces a Poisson bracket $(\cdot,\cdot)$ of degree -3 (the P-structure).
\begin{equation}
	\begin{array}{cccccc}
		x^{\mu} & \psi^{\mu} & \xi & \chi_{\mu} & \phi & p_{\mu}  \\ 
		\hline
		0 & 1 & 1 & 2 & 2 & 3
	\end{array}
\end{equation}
\begin{equation}\label{eq:D2_pb}
	(x^{\mu},p_{\nu}) = \delta^{\mu}{}_{\nu} \,, \qquad (\xi,\phi) = 1 \,, \qquad (\psi^{\mu},\chi_{\nu}) = \delta^{\mu}{}_{\nu} \,.
\end{equation}
Since the bracket is degree 3, it is antisymmetric. Other combinations of coordinates within the bracket vanishes. We will see later that this choice of QP-manifold is motivated by dimensional reduction of the M2 algebroid. Through these relations, we find the components of the antisymmetric tensor $\omega^{ab}$, and the bracket acting on general graded functions $f,g$ is given by the formula
\begin{equation}\label{eq:poisson_bracket}
	(f,g) = (-1)^{f}\del_{a}^{R}f\,\omega^{ab}\,\del_{b}g \,.
\end{equation}
Note that the definition of $\del^{R}_{a}$ is such that the combination $(-1)^{f}\del^{R}_{a}f$ is a right derivation on $f$; all other derivatives are left derivations.

The Q-structure on the manifold is a degree 1 vector field such that $Q^{2} = 0$. It is defined via a Hamiltonian function $\Theta$ of degree 4. The most general Hamiltonian we can write down is the following\footnote{A priori, one could have terms like $\alpha^{\mu}{}_{\nu}p_{\mu}\psi^{\nu}$ and $\beta^{\mu}p_{\mu}\xi$. However, we will assume that $\alpha$ is invertible, which is equivalent to saying that the anchor map associated to the algebroid is surjective. Hence, one can always perform a canonical transformation to set $\alpha^{\mu}{}_{\nu} = \delta^{\mu}{}_{\nu}$ and $\beta = 0$.}
\begin{align}
	\label{eq:D2Theta}
	\begin{split}
		\Theta &= \Theta_{0} + \Theta_{\text{flux}} \\
		&= -\psi p + \tfrac{1}{2}m \phi^{2} + J_{1}\psi\xi\phi - \tfrac{1}{2} F_{2} \psi^{2}\phi + \tfrac{1}{3!}H\psi^{3}\xi + \tfrac{1}{4!}F_{4}\psi^{4} + a\chi\phi + \tfrac{1}{2}b \chi^{2} + c \chi\psi\xi + \tfrac{1}{2}d\chi\psi^{2}  \,,
	\end{split}
\end{align}
where $\Theta_{0} = -\psi p$ and the sign $-F_{2}$ is chosen for convenience. Here $\Theta_0$ is the Hamiltonian vector field associated to the de Rham differential on $T[1]M$ while $\widetilde J_{1},F_2,F_4,H$ are all functions on $T[1]M$ i.e.~differential forms on $M$. The derivative $Q=(\Theta,\cdot)$ has the property $Q^{2}=0$ if and only if $(\Theta,\Theta)=0$, or equivalently $(\Theta_{0},\Theta_{\text{flux}}) = -\tfrac{1}{2}(\Theta_{\text{flux}},\Theta_{\text{flux}})$. Using \eqref{eq:poisson_bracket}, one can check that this implies
\begin{align}
	0 &= a=b=c=d  \,, \\
	0 &= \dr m + 2m J_{1} \label{eq:D2dRomans}  \,, \\
	0 &= \dr F_{2} - m H + J_{1}\wedge F_{2} \label{eq:D2dF2}  \,, \\
	0 &= \dr F_{4} - F_{2} \wedge H  \,, \\
	0 &= \dr H - J_{1}\wedge H  \,, \\
	0 &= \dr J_{1} \label{eq:flat_connection} \,.
\end{align}
\paragraph{Remarks.}
\begin{enumerate}
	\item For $J_1=0$, the identities arising from $Q^2=0$ correspond precisely to Bianchi identities and field equations for the fluxes of massive type IIA with the conventions of \cite{Cassani:2016ncu}.
	
	\item The field $J_{1}$ can naturally be viewed as a gauge connection for a $\bbR^{+}$ gauge transformation generated by rescalings $\xi \rightarrow e^{\lambda}\xi$, $\phi \rightarrow e^{-\lambda} \phi$. or these transformations to still satisfy \eqref{eq:D2_pb} then we also need to shift $p_{\mu} \rightarrow p_{\mu} - \del_{\mu}\lambda\xi\phi $. Putting this into \eqref{eq:D2Theta}, we see that this has the effect of transforming $J_{1} \rightarrow J_{1}+\dr\lambda$. The constraint \eqref{eq:flat_connection} then just says that this is a flat connection. Flat connections on $\bbR^{+}$ gauge bundles are always gauge trivial. That is, we can always pick a gauge where $J_{1} = 0$ globally. Hence, this is classically equivalent to the massive IIA Bianchi identities.
	
\end{enumerate}

We can relate this QP manifold with the generalised geometry of massive IIA supergravity \cite{Cassani:2016ncu}. Indeed, if we consider functions of degree $3-1=2$, we find the most general thing we can write down is
\begin{equation}
	V = v^{\mu}\chi_{\mu} - \alpha\phi + \beta_{\mu}\psi^{\mu}\xi + \tfrac{1}{2}\gamma_{\mu\nu}\psi^{\mu}\psi^{\nu}  \,,
\end{equation}
where the $-$ sign is again chosen for convenience. We can interpret this as a section of $TM\oplus \bbR \oplus T^{*}M \oplus \wedge^{2}T^{*}M$. This is the generalised tangent bundle for massive IIA supergravity, assuming the manifold $M$ has dimension $\leq3$. We can further define a bracket between 2 such sections via
\begin{equation}
	\begin{split}
		-((\Theta,V),V') &= (\mathcal{L}_{v}v')\chi \\
		& \quad \,- (\imath_{v} \dr\alpha' - \imath_{v'} (\dr\alpha - m\beta - \imath_{v} F_{2} )\phi \\
		& \quad \, +(\mathcal{L}_{v}\beta' - \imath_{v'} (\dr\beta - \imath_{v} H) )\psi\xi \\
		& \quad \, +(\mathcal{L}_{v}\gamma'  - \imath_{v'} (\dr\gamma - \beta\wedge F_{2} - \alpha H) \\
		& \qquad \qquad +\alpha'(\dr\beta - \imath_{v} H) - \beta'\wedge (\dr\alpha - m\beta - \imath_{v} F_{2}) ) \psi^{2} \\
	&= L^{F}_{V}V' \,,
	\end{split}
\end{equation}
where we have set the $J_{1}$ connection to 0. As mentioned, this is always possible and there is no global obstruction. If we had left it in the equation then all of the derivatives would be replaced with covariant derivative $D$ with respect to this $\bbR^{+}$ gauge action. We note that we find precisely (minus) the Dorfman derivative for massive IIA gravity as defined in \cite{Cassani:2016ncu}. In the following, we will always set the $J_{1}$ field to 0, knowing that we can reintroduce it by changing all derivatives into covariant derivatives.

\subsection{D4 [$P=5$ algebroid]}
Next, we consider the D4 algebroid. This will follow very similarly to above. The appropriate QP manifold $(\mathcal{M},\omega,Q)$ is given by
\begin{equation}
	\mathcal{M} = T^{*}[5]T[1]M \times \bbR[2]\times \bbR[3] \,.
\end{equation}
We can define the local homogeneous coordinates
\begin{equation}
	\begin{array}{cccccc}
		x^{\mu} & \psi^{\mu} & \xi & \phi & \chi_{\mu} & p_{\mu} \\
		\hline
		0 & 1 & 2 & 3 & 4 & 5
	\end{array}
\end{equation}
There is a natural antisymmetric Poisson bracket (P-structure) on this space given by the Poisson brackets
\begin{equation}
	(x^{\mu},p_{\nu}) = \delta^{\mu}{}_{\nu}  \,, \qquad  (\xi,\phi) = 1 \,, \qquad (\psi^{\mu},\chi_{\nu}) = \delta^{\mu}{}_{\nu} \,,
\end{equation}
which is induced by the symplectic structure
\begin{equation}
	\omega = \dr p_{\mu} \dr x^{\mu} - \dr\chi_{\mu} \dr\psi^{\mu} - \dr\xi \dr\phi \,.
\end{equation}

Once again, we can write down the Q-structure via an associated Hamiltonian function $\Theta$ of degree 5+1=6. The most general degree 6 function we can write down takes the form:
\begin{equation}
	\begin{split}
		\Theta &=\Theta_{0} + \Theta_{\text{flux}} \\
		&= -\psi p + \tfrac{1}{3!}m \xi^{3} + J_{1}\psi\xi\phi + \tfrac{1}{4} F_{2} \psi^{2}\xi^{2} + \tfrac{1}{3!}H\psi^{3}\phi + \tfrac{1}{4!}F_{4}\psi^{4}\xi + \tfrac{1}{6!}F_{6}\psi^{6} + a\chi\psi +\tfrac{1}{2}b\chi\psi^{2} \,,
	\end{split}
\end{equation}
where $\Theta_{0} = -\psi p$. Once again, the differential $Q=(\Theta,\cdot)$ satisfies $Q^{2}=0$ if and only if $(\Theta,\Theta) = 0$. We find that this is equivalent to
\begin{align}
	0&= a=b \,,\\
	0&= \dr m - 3 m J_{1} \,,\\
	0 &= \dr F_{2} - 2J_{1}\wedge F_{2} - m H \,,\\
	0 &= \dr F_{4} - J_{1}\wedge F_{4} - F_{2} \wedge H \,,\\
	0 &= \dr F_{6} - F_{4} \wedge H \,,\\
	0 &= \dr  H + J_{1} \wedge H \,,\\
	0 &= \dr  J_{1}\,.
\end{align}
\paragraph{Remarks.}
\begin{enumerate}
	\item Setting $J_{1} = 0$, we recover the Bianchi identities for massive IIA supergravity.
	
	\item $J_{1}$ is a flat $\bbR^{+}$ connection coming from the same rescaling gauge symmetry. The charges of all the fluxes under this $\bbR^{+}$ are different, however, to the D2 algebroid.
\end{enumerate}

Once again relate this structure to the generalised geometry for massive IIA supergravity of \cite{Cassani:2016ncu}. We can consider the most general homogeneous function of degree $5-1=4$ which is
\begin{equation}
	V = v\chi + \tfrac{1}{2}\alpha \xi^{2} + \beta \psi\phi + \tfrac{1}{2}\gamma\psi^{2}\xi +\tfrac{1}{4!}\delta \psi^{4} \,,
\end{equation}
which we can view as a section of $TM\oplus \bbR \oplus T^{*}M \oplus \wedge^{2}T^{*}M \oplus \wedge^{4}T^{*}M$, which is the generalised tangent bundle for massive IIA generalised geometry for $\dim M \leq 4$. Note further that $((\Theta,V),V')$ defines a bracket on the space of degree $4$ functions, i.e. generalised vector fields. Setting the gauge connection $J_{1}$ to 0 for now we find
\begin{equation}
	\begin{split}
		-((\Theta,V),V') &= (\mathcal{L}_{v}v') \chi \\
		& \quad \,  +[\mathcal{L}_{v}\alpha' - \imath_{v'} (\dr \alpha - m \beta - \imath_{v} F_{2})]\tfrac{1}{2}\xi^{2} \\
		& \quad \, +[\mathcal{L}_{v}\beta' - \imath_{v'}(\dr\beta - \imath_{v} H)]\psi\phi \\
		& \quad \,  +[\mathcal{L}_{v}\gamma' - \imath_{v'} (\dr\gamma - \alpha H -\imath_{v} F_{4} - \beta\wedge F_{2}) \\
		& \qquad \qquad \qquad - \beta' \wedge (\dr \alpha - m\beta-\imath_{v} F_{2}) + \alpha'(\dr \beta - \imath_{v} H) ] \tfrac{1}{2}\psi^{2}\xi \\
		& \quad \, +[ \mathcal{L}_{v}\delta' - \beta' \wedge(\dr \gamma - \alpha H   - \imath_{v} F_{4}- \beta\wedge F_{2}) + \gamma' \wedge(\dr \beta - \imath_{v} H) ]\tfrac{1}{4!}\psi^{4} \\
	&= L_{V}^{F}V' \,,
	\end{split}
\end{equation}
which is precisely the form of the generalised Lie derivative in \cite{Cassani:2016ncu}. Note that if we had included the $J_{1}$ in the calculation then we would get the same answer, just with $\partial \rightarrow D$, the covariant derivative associated to the $\bbR^{+}$ action.

\subsection{NS5 [$P=6$ algebroid]}

In this case we consider the QP-manifold $(\cM,\omega,Q)$ with
\be
\mathcal M= T^\star[6]T[1]M \times \mathbb{R}[3]\times \mathbb{R}[1]\times \mathbb{R}[5]\,.
\ee
In a local homogeneous in degree coordinate chart we write
\begin{equation}
	\begin{array}{ccccccc}
		x^{\mu} & \psi^{\mu} & \xi   &\zeta & \chi_{\mu}& \phi & p_{\mu}  \\ 
		\hline
		0 & 1 & 1 &3& 5 & 5 & 6
	\end{array}
\end{equation}
and define the P-structure via the nonvanishing Poisson brackets (and permutations thereof)
\begin{equation}
	(x^{\mu},p_{\nu}) = \delta^{\mu}{}_{\nu}\,, \qquad (\xi,\phi) = 1\,, \qquad (\psi^{\mu},\chi_{\nu}) = \delta^{\mu}{}_{\nu}\,,\qquad (\zeta,\zeta)=1\,,
\end{equation}
whose associated symplectic form in the convention of \cite{Arvanitakis:2018cyo} is
\be
\omega=\dr p_\mu \dr x^\mu -\tfrac{1}{2}\dr \zeta \dr \zeta - \dr\chi_\mu \dr\psi^\mu - \dr\phi \dr\xi\,.
\ee
We will see later how this choice for the NS5 is motivated via direct dimensional reduction from the M5 algebroid. One should note that, since we are in $P=6$, the Poisson bracket is antisymmetric on even coordinates but symmetric on odd.

We write down the most general Q-structure, as above, in terms of the associated Hamiltonian $\Theta$ which in this case has degree 7:
\be
\begin{split}
	\label{eq:ThetaNS5}
	\Theta &= \Theta_{0} + \Theta_{\text{flux}} \\
	&= -\psi p + J_{1}\psi\xi\phi + \tfrac{1}{2}F_{2}\psi^{2}\phi + \tfrac{1}{3!}H\psi^{3}\xi\zeta +\tfrac{1}{4!}F_{4}\psi^{4}\zeta -\tfrac{1}{6!}F_{6}\psi^{6}\xi - \tfrac{1}{7!}\widetilde{H}\psi^{7} + a\psi\xi\chi + \tfrac{1}{2}b\psi^{2}\chi \,,
\end{split}
\ee
where $\widetilde{H}$ is a 7-form which should be interpreted as the 7-form dual to $H$, the 3-form flux. We then find $Q^2=0\iff (\Theta_0,\Theta_\text{flux})=-\frac{1}{2}(\Theta_\text{flux},\Theta_\text{flux})$, which implies
\begin{equation}
\begin{split}
	0&=a=b \,,\\
	0&= \dr H +J_{1}\wedge H\,,\\
	0&=\dr  F_2 - J_{1}\wedge F_2 \label{eq:NS5:F2bianchi}\,,\\
	0&= \dr  F_4 - H\wedge F_{2}\,,\\
	0&= \dr F_6 - H\wedge F_4 + J_{1}\wedge F_6 \,,\\
	0&=\dr \widetilde{H} + \tfrac{1}{2}F_4\wedge F_{4} - F_6\wedge F_2\,,\\
	0&=\dr J_1\,.
	\end{split}
\end{equation}
\paragraph{Remarks.}
\begin{enumerate}
	\item Setting $J_{1} = 0$, we recover the Bianchi identities for \emph{massless} IIA supergravity. This is always possible as $J_{1}$ is a flat $\bbR^{+}$ connection.
	\item Note that, unlike the D2 and D4 branes, we do not get the Romans mass appearing here. Viewing the Romans mass as the flux of the D8 brane, we note that the NS5 brane only couples to the D8 brane via a D6 brane \cite{Bergshoeff:1997ak}. Our construction here, however, does not include any D6 branes since we have no $F_{8}$ flux. We therefore do not expect that the QP manifold for NS5 branes would allow for massive deformations and this is indeed what we find.
\end{enumerate}

Once again, we can relate this QP manifold to the generalised geometry of (massless) IIA supergravity \cite{Cassani:2016ncu} by considering homogeneous functions of degree $6-1=5$. These take the general form
\begin{equation}
	V = -v\chi +\alpha\phi + \beta\psi\xi\zeta +\tfrac{1}{2}\gamma\psi^{2}\zeta -\tfrac{1}{4!} \delta \psi^{4}\xi +\tfrac{1}{5!}\epsilon\psi^{5}\,.
\end{equation}
This can be identified with a section of $TM\oplus T^{*}M \oplus \bbR\oplus \wedge^{2}T^{*}M \oplus \wedge^{4}T^{*}M \oplus \wedge^{5}T^{*}M$, which is the generalised tangent bundle for IIA supergravity provided $\dim M \leq 5$. We can then define a bracket between any two such functions via
\begin{equation}
	\begin{split}
		-((\Theta,V,V') &= -\mathcal{L}_{v}v' \chi \\
		& \quad \, +[\mathcal{L}_{v}\alpha' - \imath_{v'}(\dr\alpha - \imath_{v} F_{2})]\psi\xi\phi \\
		& \quad \, +[\mathcal{L}_{v}\beta' - \imath_{v'} (\dr\beta - \imath_{v} H)] \psi\xi\zeta \\
		& \quad \, +[\mathcal{L}_{v}\gamma' - \imath_{v'} (\dr\gamma - \alpha H - \beta\wedge F_{2} - \imath_{v} F_{4} ) \\
		& \qquad \qquad \;\;\, - \beta'\wedge (\dr\alpha - \imath_{v} F_{2}) +\alpha'(\dr\beta - \imath_{v} H)] \tfrac{1}{2}\psi^{2}\zeta \\
		&\quad \, - [\mathcal{L}_{v}\delta' -\imath_{v'}(\dr\delta - \gamma\wedge H - \beta\wedge F_{4}) \\
		& \qquad \qquad \;\;- \beta'\wedge(\dr\gamma - \alpha H - \beta\wedge F_{2} - \imath_{v} F_{4}) + \gamma'\wedge(\dr\beta - \imath_{v} H)] \tfrac{1}{4!}\psi^{4}\xi \\
		& \quad \, +[\mathcal{L}_{v}\epsilon' +\alpha'(\dr\delta - \gamma\wedge H - \beta\wedge F_{4}) \\
		& \qquad \qquad \;\; - \gamma'\wedge(\dr\gamma - \alpha H - \beta\wedge F_{2} - \imath_{v} F_{4}) + \delta' \wedge (\dr\alpha - \imath_{v} F_{2})  ]\tfrac{1}{5!}\psi^{5}\\
	&= L^{F}_{V}V'\,.
	\end{split}
\end{equation}
We see that we precisely get the form of the Dorfman derivative as in \cite{Cassani:2016ncu}.

\section{IIA/M-theory duality}\label{sec:M/IIA_duality}
\subsection{M5 to NS5}
We start with the M5 algebroid of \cite{Arvanitakis:2018cyo}. This is given by the graded manifold $\cM_\text{M5}=T^\star[6]T[1]N\times \mathbb{R}[3]$. Here we choose $N$ to be a principal $S^1$-bundle over a manifold $M$:
$$
S^1\into N\onto M\,.
$$
The fibre will be playing the role of the M-theory circle. We will be working in coordinates
\begin{equation}
	\begin{array}{cccccccccc}
		x^{\mu} & y& \psi^{\mu}  & \xi & \zeta& \phi & \chi_{\mu} & p_{\mu} &\pi \\
		\hline
		0 & 0& 1 & 1 & 3 & 5 & 5 & 6&6
	\end{array}
\end{equation}
on $\cM_\text{M5}$ that are adapted to a local trivialisation of $N$, such that the index-free coordinates $\{y,\xi,\phi,\pi\}$ coordinatise $T^\star[6]T[1]S^1$; $\phi$ and $\xi$ in particular are associated to the fundamental vector field and Maurer-Cartan 1-form on $S^1$ respectively. We will also need a connection for the $S^1$ bundle. This is equivalently a certain 1-form on $N$, which we can identify with the degree 1 function $\mathcal A$ on $T[1]N$ that takes the form
\be
\label{connection:in:coords}
\mathcal A=\xi + A_\mu(x)\psi^\mu\,,
\ee
for $A_\mu$ a locally-defined gauge potential on the base $M$.

We make the obvious choice for the P-structure that accords with the one we made for the NS5. The Q-structure of the M5 algebroid depends on two forms $G_4$ and $G_7$ which obey the familiar M-theory Bianchi identities $\dr G_4=0$ and $\dr G_7+\frac{1}{2} G_4 G_4=0$. It can be checked that these are equivalent to nilpotency of the Q-structure $Q_\text{M5}\equiv (\Theta_\text{M5},\bullet)$ that arises from
\be
\label{eq:thetaM5}
\Theta_\text{M5}=-\psi^\mu p_\mu -\pi \xi + G_7 + G_4\zeta\,.
\ee
The forms $G_4$ and $G_7$ can be expanded in terms of basic forms $L_n$ and the connection $\mathcal A$ as follows:
\be
G_7= L_7 + \mathcal A L_6\,,\qquad G_4= L_4 + \mathcal A L_3\,,
\ee
where here and henceforth we identify basic forms on $N$ with the forms on $M$ whose pullbacks produce basic forms on $N$.

The M-theory/type IIA duality will be realised via symplectic reduction with respect to the Hamiltonian lift of the canonical $S^1$ action on $N$. (For this we follow the recent treatment \cite{Arvanitakis:2021lwo}, also in a graded symplectic context). While this gives rise to a $T[1]S^1$ action on $T[1]N$, we will only be gauging the degree zero subgroup (i.e.~$S^1$), which will have the effect of retaining the `extra' coordinates $\phi,\,\xi$. This Hamiltonian lift of the $S^1$ action is generated by $\pi$. Since $\pi$ is of degree 6, the associated Hamiltonian vector field $(\pi,\bullet)$ is degree-preserving, as expected. The Q-structure will reduce in this case if and only if $Q_\text{M5} \pi= 0\mod \pi$, which states the fluxes $G_4,G_7$ must be $S^1$-invariant.

Instead of reducing the Q-structure of \eqref{eq:thetaM5} immediately, we will first perform a symplectomorphism that produces a new Q-structure $\bar Q$, where the locally-defined connection 1-form $A_\mu$ (appearing through $\mathcal A$) is traded in favour of its field strength 2-form which is globally-defined on $M$. This way the fluxes appearing in the Q-structure will be manifestly basic forms on $N$, and will thus descend to forms on the base $M$.\footnote{The necessity of doing a symplectomorphism before the symplectic reduction was also observed in \cite{Arvanitakis:2021lwo}.} The symplectomorphism is in this case generated by the degree 6 function $-\phi A_\mu \psi^\mu$. This acts trivially on all coordinates except for
\be
\label{eq:M5toNS5symplecto}
\xi\to \xi - A_\mu \psi^\mu\,,\qquad p_\mu \to p_\mu -\phi \pd_\mu A_\nu \psi^\nu\,, \qquad \chi_\mu \to \chi_\mu - \phi A_\mu\,.
\ee
This map transforms the Q-structure \eqref{eq:thetaM5} into
\be
\label{eq:barthetaM5def}
\Theta_\text{M5}\to \bar \Theta\equiv -\psi^\mu p_\mu -\pi \xi +L_7+\xi L_6 + (L_4 + \xi L_3) - \phi F\,,
\ee
where $F$ is the field strength of the connection in the standard convention $F_{\mu\nu}= 2\pd_{[\mu} A_{\nu]}$. Since $\mathcal A\to \xi$ we observe that the effect of the symplectomorphism is to identify the new $\xi$ with the connection $\mathcal A$, from which we infer that $\phi$ is identified with the fundamental vector field generating the $S^1$ action.

It is now easy to calculate how the Q-structure of \eqref{eq:barthetaM5def} reduces under the symplectic reduction generated by $\pi$. We evidently get the Q-structure of an NS5 algebroid with the identifications
\be
L_7=-\widetilde H\,, \quad L_6=-F_6\,, \quad  L_4=F_4\,,\quad L_3=-H\,,\quad F=-F_2\,,\quad 0=J_1\,.
\ee
The geometric considerations that led to the introduction of the symplectomorphism \eqref{eq:M5toNS5symplecto} have also produced the correct identification of the type IIA Ramond-Ramond 2-form flux with the curvature of the connection associated to the M-theory circle bundle.

\subsection{M2 to D2}
We parallel the derivation we gave for the M5 reduction above. The QP manifold associated to the M2 brane is $\cM_\text{M2}=T^\star[3]T[1]N$ \cite{Kokenyesi:2018ynq}. We again choose $N$ to be a principal $S^1$-bundle over a manifold $M$. We will be working in coordinates
\begin{equation}
	\begin{array}{ccccccccc}
		x^{\mu} & y& \psi^{\mu}  & \xi & \phi & \chi_{\mu} & p_{\mu} &\pi \\
		\hline
		0 & 0& 1 & 1  & 2 & 2 & 3&3
	\end{array}
\end{equation}
The Q-structure in this case is determined by a 4-form that we expand in horizontal forms via a connection $\mathcal A$ for the $S^1$-bundle, so $G_4=L_4 + \mathcal A L_3$. Then the Q-structure in these coordinates can be expressed via its Hamiltonian which takes the form
\be
\Theta_\text{M2}= -\psi^\mu p_\mu-\pi \xi + L_4 + \mathcal A L_3\,.
\ee

We again perform a canonical transformation generated, in this case, by the degree +3 function $\phi A_\mu(x)\psi^\mu$. This in particular has the effect of replacing
$\mathcal A\to \xi$
(c.f.~\eqref{connection:in:coords}) and
\be
\Theta_\text{M2}\to \bar \Theta= -\psi^\mu p_\mu-\pi \xi + L_4 + \xi L_3-\phi F\,,
\ee
for $F\equiv\dr A$. Assuming the original 4-form $G_4$ was $S^1$-invariant, we find that all three forms $L_4,L_3,F$ are basic, which implies that $\bar Q \pi=0.$ Therefore we can immediately perform a symplectic reduction with respect to the $S^1$ action generated by $\pi$. The result is clearly the graded manifold $\mathcal M_\text{D2}=T^\star[3]T[1]M\times \mathbb R[1]\times \mathbb R[2]$ for the D2-brane given above, with the P structure as in that previous subsection, and with Q-structure determined by the fluxes
\be
F_4=L_4\,,\quad H=-L_3\,,\quad F_2=-F\,,\quad F_1=0\,,\quad F_0=0\,.
\ee

We find in particular that this reduction of the M2 algebroid to a D2 algebroid always generates a D2 algebroid with vanishing Romans mass parameter $F_0$. This is in accord with the known lore that the Romans massive IIA theory cannot be obtained through an ordinary Scherk-Schwarz compactification of M-theory. It \emph{can} be found by dualising IIB with linear $F_1$ flux, equivalently as a certain compactification of F-theory \cite{Hull:1998vy}, or as a \emph{generalised Scherk-Schwarz compactification} of exceptional field theory \cite{Ciceri:2016dmd} that involves an explicit linear winding/dual coordinate dependence in the frame field.\footnote{In double field theory, Romans IIA theory arises by giving the Ramond-Ramond fields a linear dependence on the dual coordinates \cite{Hohm:2011cp}.} 

\section{Applications}\label{sec:applications}

\subsection{AKSZ models}
Each QP manifold has an associated topological field theory given by the AKSZ construction \cite{Alexandrov:1995kv}. For certain choices of QP manifolds, the associated action functionals turn out to be related to the Wess-Zumino couplings between strings, M5- and D3-branes and the relevant background fluxes \cite{Ikeda:2002qx,Roytenberg:2006qz,Arvanitakis:2018cyo}. Here we will establish a similar relation for the putative `D2-brane' algebroid and the `NS5-brane' algebroid, and thereby partly justify the appellation. We do not do the `D4-brane' algebroid explicitly here as it will follow very similarly to the D2 algebroid. We will also set the $J_{1}$ connection to 0 in the following and will follow the work of \cite{Arvanitakis:2018cyo} closely.

\subsubsection{D2 Brane}

The AKSZ construction builds a topological theory on $\maps(T[1]\Sigma \rightarrow \mathcal{M})$ where $\Sigma$ is a dimension $P+1$ manifold if $\mathcal{M}$ is a QP manifold of degree $P$. For a degree $P=3$ QP manifold, the AKSZ construction therefore requires $\dim \Sigma_{4} = 4$ and we will assume that it has boundary
\begin{equation}
	\del \Sigma_{4} = W_{3} \,,
\end{equation}
where we can interpret $W_{3}$ as the worldvolume of the D2 brane. The AKSZ action is then given by
\begin{align}
	S &= S_{\text{bulk}} +S_\text{boundary} \,, \qquad
	S_{\text{bulk}} = \int_{\Sigma_{4}} \vartheta - \Theta \,,
\end{align}
where $\vartheta$ is a symplectic potential, i.e. $\dr \vartheta = \omega$, and the boundary action is chosen in conjunction with consistent boundary conditions on the fields such that the resulting equations of motion define stationary points of $S_{\text{bulk}}$. We will see that $S_{\text{boundary}}$ is the WZ term for the D2 brane as found in e.g. \cite{Green:1996bh}.

On any QP manifold, one has a degree counting vector field, or Euler vector field, $\varepsilon$ such that, for any homogeneous coordinate $z^{a}$ one has
\begin{equation}
    \varepsilon(z^{a}) = (\mathrm{deg}z^{a})z^{a} \,.
\end{equation}
We can use this to define a global $\vartheta$ by $\vartheta_\text{{Euler}} = \tfrac{1}{P}\imath_{\varepsilon}\omega$, where $P$ is the degree of the symplectic form (see e.g. \cite[Appendix A.1]{Arvanitakis:2018cyo} for more details). However, this does not turn out to be the correct choice to give the D2 brane WZ term. It is possible to show that any two symplectic potentials differ by the exterior derivative acting on a function of degree $P$. It turns out that the `good' choice for $\vartheta$ is given by
\begin{equation}
	\begin{split}
    \vartheta_{\text{good}} &= \vartheta_{\text{Euler}} + \tfrac{2}{3}\dr(\xi\phi) \\
    &= p\dr x - \tfrac{1}{3}\psi\dr\chi - \tfrac{2}{3}\chi\dr\psi - \xi\dr\phi \,. 
    \end{split}\label{eq:vartheta_good_D2}
\end{equation}
The choice of homogeneous function $\xi\phi$ is globally well defined. Indeed, writing $\mathcal{M}=T^{*}[3](T[1]M\times \bbR[1])$, the function $\xi\phi$ is the Hamiltonian that generates the $\bbR^{+}$ rescaling of $\bbR[1]$. That is, it generates the $\bbR^{+}$ rescaling for which $J_{1}$ is a connection. This provides a remarkable interpretation for the mysterious field $J_{1}$ as being necessary for the WZ terms and suggests that, by leaving $J_{1}$ explicit in the calculation, one could do the AKSZ construction in a $\vartheta$-agnostic way. For simplicity, we shall not present that argument here, and proceed without $J_{1}$ and with $\vartheta = \vartheta_{\text{good}}$ as given in \eqref{eq:vartheta_good_D2}.

The bulk action is given by
\begin{align}
    \begin{split}
        S_{\text{bulk}} &= \int_{\Sigma_{4}} p\dr x - \tfrac{1}{3}\psi\dr\chi - \tfrac{2}{3}\chi\dr\psi - \xi\dr\phi \\
        & \qquad \qquad - \left( -\psi p + \tfrac{1}{2}m \phi^{2} - \tfrac{1}{2} F_{2} \psi^{2}\phi + \tfrac{1}{3!}H\psi^{3}\xi + \tfrac{1}{4!}F_{4}\psi^{4} \right) \,,
    \end{split}
\end{align}
where we interpret $\psi,\xi$ to be 1-forms on $\Sigma_{4}$, $\chi,\phi$ to be 2-forms, and $p$ to be 3-forms. We consider the boundary terms arising from a variation of $S_{\text{bulk}}$, which is given by
\begin{equation}\label{eq:boundary_variation_D2}
    I = \int_{W_{3}} -p\delta x + \tfrac{1}{3}\psi\delta\chi - \tfrac{2}{3}\chi\delta \psi + \xi\delta\phi\,, \quad \text{where} \quad \delta S_{\text{bulk}} = I+ \int_{\Sigma_{4}}(\text{e.o.m})\,.
\end{equation}
It is always consistent to substitute fields for their own equations of motion. Looking at the equations of motion for $p,\psi,\xi,\phi$ we find respectively
\begin{align}
    \psi^{\mu} &= \dr x^{\mu} \label{eq:D2_psi_eom}\,, \\
    p_{\mu} &= \dr\chi_{\mu} - F_{2 \mu}\psi\phi + \tfrac{1}{2}H_{\mu}\psi^{2}\xi + \tfrac{1}{3!}F_{4\mu}\psi^{3} \label{eq:D2_p_eom} \,,\\
    \dr\phi &= \tfrac{1}{3!}H\psi^{3} \label{eq:D2_phi_eom}\,, \\
    \dr\xi &= \tfrac{1}{2}F_{2}\psi^{2} - m\phi \label{eq:D2_xi_eom}\,,
\end{align}
where e.g. $H_{\mu}\psi^{2}$ is shorthand for $H_{\mu\nu\rho}\psi^{\nu}\psi^{\rho}$. Note that substituting in \eqref{eq:D2_psi_eom} means that we can interpret all functions $F_{p},H$ as pull-backs of differential forms on the target space $M$. Substituting \eqref{eq:D2_psi_eom} and \eqref{eq:D2_p_eom} into \eqref{eq:boundary_variation_D2}, we find
\begin{align}
    \begin{split}
    I &= \int_{W_{3}} -\dr\chi\delta x + (\imath_{\delta x}F_{2})\phi - (\imath_{\delta x}H)\xi - (\imath_{\delta x} F_{4}) + \tfrac{1}{3}\dr x \delta \chi - \tfrac{2}{3}\chi\delta \dr x + \xi\delta \phi \\
    &= \int_{W_{3}} \delta(\tfrac{1}{3}\chi\dr x) + (\imath_{\delta x}F_{2})\phi - (\imath_{\delta x}H)\xi - (\imath_{\delta x} F_{4}) + \xi\delta \phi \,.
    \end{split} \label{eq:boundary_varyation_D2_b}
\end{align}
We can remove the first term by setting $\chi|_{W_{3}} = 0$, which is consistent with the equations of motion \eqref{eq:D2_psi_eom}-\eqref{eq:D2_xi_eom}. Next we use the following results which follow from the solution to the Bianchi identities $H=\dr B$, $F_{p} = \dr C_{p-1} - H\wedge C_{p-3} + m\,e^{B}$.
\begin{equation}
	\begin{split}
    \imath_{\delta x} H &= \delta B - \dr(\imath_{\delta x} B) \,, \\
    \imath_{\delta x} F_{2} &= \delta C_{1} - \dr(\imath_{\delta x}C_{1}) + m\,\imath_{\delta x} B \,, \\
    \imath_{\delta x} F_{4} &= \delta C_{3} - \dr(\imath_{\delta x}C_{3})  - (\delta B - \dr(\imath_{\delta x} B))\wedge C_{1} + H\wedge \imath_{\delta x}C_{1} + m\,\imath_{\delta x} B\wedge B \,.
    \end{split}
\end{equation}
Substituting these into \eqref{eq:boundary_varyation_D2_b}, performing some integration by parts and using the equations of motion, we find this becomes
\begin{equation}\label{eq:boundary_variation_D2_c}
    I = \int_{W_{3}} -\delta C_{3} + \delta B\wedge C_{1} - \delta B \wedge \xi + \delta C_{1} \wedge \phi + \xi \wedge \delta \phi \,,
\end{equation}
where we have reintroduced the $\wedge$ to make the interpretation of these objects as (pull-backs of) differential forms on $M$ explicit. Finally, we can solve the equations of motion for $\phi, \xi$ given in \eqref{eq:D2_phi_eom} and \eqref{eq:D2_xi_eom} by
\begin{equation}
	\begin{split}
    \phi &= B- \tfrac{\alpha'}{2\pi}F \,, \\
    \xi &= C_{1} + m\tfrac{\alpha'}{2\pi} A + X \,,
    \end{split}
\end{equation}
where $\dr F = \dr X = 0$ and $\dr A = F$. The factors of $\alpha'$ are introduced for convenience later. We will see that it is natural to associate $A$ with the Born-Infeld 1-form and $F$ its field strength. Indeed, this identification will match the WZ term given in \cite{Green:1996bh}. Furthermore, this combination of $B,F$ is invariant under NS-NS gauge transformations \cite{Lozano:1997ee}, just as $\phi$ should be. Substituting these into \eqref{eq:boundary_variation_D2_c}, we find
\begin{equation}
    I = \int_{W_{3}}-\delta \left( C_{3} + C_{1}\wedge(\tfrac{\alpha'}{2\pi}F - B) + \tfrac{1}{2}(\tfrac{\alpha'}{2\pi})^{2}m\,A\wedge F\right) - X\delta F \,.
\end{equation}
Imposing the condition that $X|_{W_{3}} = 0$ we see that $I = -\delta I_{WZ}$ \cite{Green:1996bh}.

We now choose $S_{\text{boundary}}$ so that the equations of motion \eqref{eq:D2_psi_eom} - \eqref{eq:D2_xi_eom} are stationary points of the full action $S$. Hence, the variation of $\S_{\text{boundary}}$ must cancel $I$ above, and we see that we need
\begin{equation}
    S_{\text{boundary}} = I_{WZ} \,,
\end{equation}
where here $I_{WZ}$ stands the Wess-Zumino action of the D2 brane \cite{Green:1996bh}. This then implies, as desired
\begin{equation}
    \delta S = \delta S_{\text{bulk}} + \delta S_{\text{boundary}} = \int_{\Sigma_{4}}(\text{e.o.m}) \,.
\end{equation}

\subsubsection{NS5 brane}

We will now perform the same analysis for the NS5 brane. We won't perform the analysis for the D4 brane since it follows very similarly to the D2 brane above. We once again build a theory on $\maps({T[1]\Sigma_{7} \rightarrow \mathcal{M}})$ where $\Sigma_{7}$ is a 7 dimensional manifold such that the boundary
\begin{equation}
    \del\Sigma_{7} = W_{6} \,,
\end{equation}
is interpreted as the worldvolume of the NS5 brane. We take the AKSZ action given by
\begin{equation}
    S = S_{\text{bulk}} + S_{\text{boundary}} \qquad S_{\text{bulk}} = \int_{\Sigma_{7}}\vartheta+\Theta \,.
\end{equation}
Note that the relative sign in the bulk action is different to that chosen for the D2 brane. Ultimately, the relative sign is not important for the AKSZ construction and $+$ is only chosen here for convenience. We take the following `good' choice of symplectic potential.
\begin{equation}
	\begin{split}
    \vartheta_{\text{good}} &= \vartheta_{\text{Euler}} + \tfrac{1}{3}\dr(\xi\phi) \\
    &=p\dr x - \tfrac{1}{6}\psi\dr\chi - \tfrac{5}{6}\chi\dr\psi - \tfrac{1}{2}\xi\dr\phi - \tfrac{1}{2}\phi\dr\xi - \tfrac{1}{2}\zeta\dr\zeta \,.
   	\end{split}
\end{equation}
Note that, once again, the good choice of $\vartheta$ differs from the canonical choice by a global function which is the Hamiltonian generating the $\bbR^{+}$ action for which $J_{1}$ is a gauge potential.

The bulk action is
\begin{align}
    \begin{split}
        S_{\text{bulk}} &= \int_{\Sigma_{7}} p\dr x - \tfrac{1}{6}\psi\dr\chi - \tfrac{5}{6}\chi\dr\psi - \tfrac{1}{2}\xi\dr\phi - \tfrac{1}{2}\phi\dr\xi - \tfrac{1}{2}\zeta\dr\zeta \\
        & \qquad -\psi p + \tfrac{1}{2}F_{2}\psi^{2}\phi + \tfrac{1}{3!}H\psi^{3}\xi\zeta +\tfrac{1}{4!}F_{4}\psi^{4}\zeta -\tfrac{1}{6!}F_{6}\psi^{6}\xi - \tfrac{1}{7!}\widetilde{H}\psi^{7} \,.
    \end{split}
\end{align}
We vary the action and consider the boundary terms which arise.
\begin{equation}\label{eq:NS5_boundary_variation_a}
    I = \int_{W_{6}} p\delta x + \tfrac{1}{6} \psi \delta \chi + \tfrac{5}{6}\chi\delta \psi + \tfrac{1}{2}\xi\delta \phi + \tfrac{1}{2}\phi\delta \xi + \tfrac{1}{2}\zeta \delta \zeta
\end{equation}
Once again, it is always consistent to substitute fields for their own equation of motion. Looking at the equations of motion for $p,\psi,\phi,\zeta,\xi$, we find respectively
\begin{align}
    \psi^{\mu} &= \dr x^{\mu} \label{eq:NS5_psi_eom} \,, \\
    p_{\mu} &= -\dr\chi_{\mu} + F_{2\mu}\psi\phi + \tfrac{1}{2}H_{\mu}\psi^{2}\xi\zeta + \tfrac{1}{3!}F_{4\mu}\psi^{3}\zeta - \tfrac{1}{5!}F_{6\mu}\psi^{5}\xi - \tfrac{1}{6!}\widetilde{H}_{\mu}\psi^{6} \label{eq:NS5_p_eom} \,, \\
    \dr\xi &= \tfrac{1}{2}F_{2}\psi^{2} \label{eq:NS5_xi_eom} \,, \\
    \dr\zeta &= \tfrac{1}{3!}H\psi^{3}\xi + \tfrac{1}{4!}F_{4}\psi^{4} \label{eq:NS5_zeta_eom}	\,, \\
    \dr\phi &= -\tfrac{1}{3!}H\psi^{3}\zeta - \tfrac{1}{6!}F_{6}\psi^{6} \,. \label{eq:NS5_phi_eom}
\end{align}
Substituting \eqref{eq:NS5_psi_eom} and \eqref{eq:NS5_p_eom} into \eqref{eq:NS5_boundary_variation_a}, we find
\begin{align}
    \begin{split}\label{eq:NS5_boundary_variation_b}
    I &= \int_{W_{6}} -\delta(\tfrac{1}{6}\chi\dr x) + (\imath_{\delta x}F_{2})\phi + (\imath_{\delta x}H)\xi\zeta + (\imath_{\delta x} F_{4})\zeta - (\imath_{\delta x}F_{6})\xi -(\imath_{\delta x}\widetilde{H}) \\
    &\qquad \qquad + \tfrac{1}{2}\xi\delta \phi +\tfrac{1}{2}\phi\delta \xi + \tfrac{1}{2}\zeta\delta\zeta
    \end{split}
\end{align}
We can remove the first term with the boundary condition $\chi|_{W_{6}} = 0$, which is consistent with the equations of motion. We then use the following results which follow from the solutions to the Bianchi identities $H=\dr B$, $F_{p} = \dr C_{p-1} - H\wedge C_{p-3}$, $\widetilde{H} = \dr \widetilde{B} + \tfrac{1}{2}(-F_{2}C_{5} + F_{4}C_{3} - F_{6}C_{1})$.
\begin{equation}
    \begin{split}
    \imath_{\delta x} H &= \delta B - \dr(\imath_{\delta x} B)	\,, \\
    \imath_{\delta x} F_{2} &= \delta C_{1} - \dr(\imath_{\delta x}C_{1})	\,, \\
    \imath_{\delta x} F_{4} &= \delta C_{3} - \dr(\imath_{\delta x}C_{3})  - (\delta B - \dr(\imath_{\delta x} B))\wedge C_{1} + H\wedge \imath_{\delta x}C_{1}	\,, \\
    \imath_{\delta x} F_{6} &= \delta C_{5} - \dr(\imath_{\delta x}C_{5})  - (\delta B - \dr(\imath_{\delta x} B))\wedge C_{3} + H\wedge \imath_{\delta x}C_{3}	\,, \\
    \imath_{\delta x}\widetilde{H} &= \delta \widetilde{B} - \dr(\imath_{\delta x}\widetilde{B}) -\tfrac{1}{2}(\delta C_{1} - \dr(\imath_{\delta x} C_{1}) \wedge C_{5} - \tfrac{1}{2}F_{2}\wedge \imath_{\delta x}C_{5} \\
    & \qquad +\tfrac{1}{2}(\delta C_{3} - \dr(\imath_{\delta x}C_{3})  - (\delta B - \dr(\imath_{\delta x} B))\wedge C_{1} + H\wedge \imath_{\delta x}C_{1})\wedge C_{3} + \tfrac{1}{2}F_{4}\wedge \imath_{\delta x}C_{3} \\
   &\qquad -\tfrac{1}{2}(\delta C_{5} - \dr(\imath_{\delta x}C_{5})  - (\delta B - \dr(\imath_{\delta x} B))\wedge C_{3} + H\wedge \imath_{\delta x}C_{3})\wedge C_{1} - \tfrac{1}{2}F_{6} \wedge \imath_{\delta x}C_{1} \,.
    \end{split}
\end{equation}
Substituting these into \eqref{eq:NS5_boundary_variation_b} we find
\begin{align}
    \begin{split}\label{eq:NS5_boundary_variation_c}
        I &= \int_{W_{6}} \delta C_{1} \phi + \delta B \xi\zeta +\delta C_{3}\zeta - \delta BC_{1}\zeta - \delta C_{5}\xi + \delta BC_{3}\xi - \delta \widetilde{B} + \tfrac{1}{2}\delta C_{1}C_{5} \\
        & \qquad - \tfrac{1}{2}\delta C_{3}C_{3} + \delta BC_{1}C_{3} + \tfrac{1}{2}\delta C_{5}C_{1} + \tfrac{1}{2}\xi\delta \phi + \tfrac{1}{2}\phi\delta \xi + \tfrac{1}{2}\zeta\delta\zeta \,.
    \end{split}
\end{align}
We can then solve the equations of motion for $\xi,\zeta,\phi$ given in \eqref{eq:NS5_xi_eom}-\eqref{eq:NS5_phi_eom} by
\begin{equation}
	\begin{split}
    \xi &= \mathcal{G}_{1} = C_{1} + \dr c_{0} \,, \\
    \zeta &= \mathcal{G}_{3} = C_{3}- Hc_{0} + \dr c_{2} \,, \\
    -\phi &= \mathcal{G}_{5} = C_{5} - H c_{2} + \dr c_{4} \,,
    \end{split}
\end{equation}
where we have used the same notation as \cite{deBoer:2012ma}, and where $c_{p}$ are locally differential forms which transform in such a way that $\mathcal{G}_{p'}$ are gauge invariant. Substituting these into \eqref{eq:NS5_boundary_variation_c} we find
\begin{equation}
    I = \int_{W_{3}} - \delta \left( \widetilde{B} + \tfrac{1}{2}(-\mathcal{G}_{5}C_{1} + \mathcal{G}_{3}C_{3} - \mathcal{G}_{1}C_{5}) \right)+ \ldots = -\delta I_{WZ} \,,
\end{equation}
where the $\dots$ corresponds to terms quadratic in the $c_{p}$ which vanish when one enforces a duality constraint relating $c_{0}$ and $c_{4}$, and a self-duality constraint on $c_{2}$. This matches the constraints originally found in \cite{Bergshoeff:2011zk} for these worldvolume fields. Hence, the boundary term is precisely the variation of (minus) the WZ term for the NS5 brane \cite{deBoer:2012ma}. The boundary action $S_{\text{boundary}}$ must be chosen to cancel this term and hence we need
\begin{equation}
    S_{\text{boundary}} = I_{WZ} \,,
\end{equation}
where here $I_{WZ}$ stands the Wess-Zumino action of the NS5 brane, e.g. \cite{deBoer:2012ma}.

\subsection{The (topologically massive) D2 brane phase space}
We first consider the case where all fluxes except the Romans mass $m$ have been switched off. Then the Hamiltonian $\Theta$ of \eqref{eq:D2Theta} associated to the Q-structure $Q_\cM$ on $\cM=T^\star[3]T[1]M\times \mathbb R[1]\times \mathbb R[2]$ simplifies to 
\be
\Theta=-\psi^\mu p_\mu +\tfrac{1}{2}m\phi^2\,,
\ee
with $m$ constant. 

The construction of \cite{Arvanitakis:2021wkt} produces a degree-zero Poisson bracket on the subspace of
\be
\maps(T[1] \Sigma_2 \to \cM) \,,
\ee
defined by the zero locus of the vector field $\dr_{T[1]\Sigma_2}-Q_{\cM}$. This induces an ordinary Poisson bracket on the degree-zero subspace of that zero locus, which we will now identify as the phase space of a D2 brane on a Romans massive background (the brane worldvolume being $\mathbb R\times \Sigma_2$). We will focus on the novelty of the D2 brane versus the case of the M2 brane (which was already discussed in that reference), which is the pair of $\mathbb R[1]\times \mathbb R[2]$ coordinates $\xi$ and $\phi$ and their contribution to the brane phase space.

The coordinates $\xi$ and $\phi$ on the target space $\cM$ respectively give rise to degree 1 and 3 superfields $\bm{\xi}$ and $\bm{\phi}$, which are some of the components of a map $T[1] \Sigma_2 \to \cM$. These are expanded in component fields as
\be
\begin{split}
	\bm{\phi}=\tfrac{1}{2}\phi_{\alpha_1\alpha_2}(\sigma)\varepsilon^{\alpha_1\alpha_2} \dr \sigma^1 \dr \sigma^2+\ldots\equiv \Phi(\sigma)\dr \sigma^1 \dr \sigma^2+\ldots\,,\qquad
	\bm{\xi}= \xi_\alpha(\sigma) \dr \sigma^\alpha +\ldots\,,
\end{split}
\ee
where $\sigma^\alpha=(\sigma^1,\sigma^2)$ are coordinates on $\Sigma_2$, $\varepsilon^{\alpha\beta}$ is the invariant Levi-Civita tensor density with $\varepsilon^{12}=1$, $\deg \dr \sigma^\alpha=1$, and the ellipses denote component fields of non-zero degree (``ghosts''). The brane phase space construction then gives a Poisson bracket formula for smeared currents such as
\be
-\langle \xi|\epsilon \rangle\equiv \int_{T[1]\Sigma_2} \bm{\xi} \epsilon \,,
\ee
for $\epsilon$ any form on $\Sigma_2$. This formula takes the form
\be
\label{eq:PBformula}
\big\{\langle f|\epsilon\rangle\,, \langle g|\eta\rangle\big\}= \pm \langle (f, Q_{\cM}g)|\epsilon \eta\rangle \pm \langle (f,g)|\epsilon \dr_{T[1]\Sigma_2}\eta\rangle \,,
\ee
for $f,g$ any functions on $\cM$. We refer to \cite{Arvanitakis:2021wkt} for the signs and more details.

We see from \eqref{eq:PBformula} that the Romans mass will appear in the Poisson brackets of $\Phi(\sigma)$ and $\xi_\alpha(\sigma)$. Indeed we derive the non-vanishing Poisson brackets (where $\varepsilon_{12}=1$)
\be
\label{eq:D2PBs}
\{\Phi(\sigma),\xi_\alpha(\tau)\}= \frac{\pd}{\pd\tau^\alpha}\delta^2(\sigma-\tau)\,,\qquad \{ \xi_\alpha(\sigma),\xi_\beta(\tau)\}= m \varepsilon_{\alpha\beta} \delta^2(\sigma-\tau)\,,
\ee
with the currents associated to the other variables ($x,\psi,\chi,p$) giving the standard brane position/momentum Poisson brackets. 

The D2-brane worldvolume on a Romans massive background carries a gauge field which evolves according to topologically massive electrodynamics \cite{Deser:1981wh} (coupled to the worldvolume scalar fields determining the brane embedding in spacetime). The phase space of this theory is given in terms of the spatial vector potential $A_\alpha(\sigma)$ and its conjugate momentum $D^\alpha(\sigma)$, with gauge transformations generated by the Gauss law constraint $G$:
\be
\{A_\alpha(\sigma), D^\beta(\tau)\}=\delta^\beta_\alpha \delta^2(\sigma-\tau)\,, \qquad G\equiv \pd_\alpha D^\alpha - \tfrac{1}{2}m \varepsilon^{\alpha\beta}\pd_\alpha A_\beta\,.
\ee
The dynamics of the theory is given by the Hamiltonian 
\be
H=\int d^2\sigma\; (E^2+B^2)\,,\qquad B\equiv \varepsilon^{\alpha\beta}\pd_\alpha A_\beta\,,\quad E^\alpha\equiv D^\alpha + \tfrac{1}{2}m\varepsilon^{\alpha \beta}A_\beta\,.
\ee
This Hamiltonian is manifestly gauge-invariant, as the variables $B$ and $E^\alpha$ Poisson-commute with the Gauss law constraint $G$.

The Poisson brackets \eqref{eq:D2PBs} obtained in the QP manifold picture are precisely the Poisson brackets of the gauge-invariant variables
\be
\{E^\alpha(\sigma),E^\beta(\tau)\}=m\varepsilon^{\alpha\beta}\delta^2(\sigma-\tau)\,,\quad \{E^\alpha(\sigma),B(\tau)\}=\varepsilon^{\alpha\beta}\frac{\pd}{\pd\tau^\beta}\delta^2(\sigma-\tau)\,,
\ee
if we identify $E^\alpha=\varepsilon^{\alpha\beta}\xi_\beta$ and $\Phi=-B$. Furthermore, the zero-locus condition ($\dr_{T[1]\Sigma_2}-Q_{\cM}=0$) implies
$\dr_{T[1]\Sigma_2}\bm{\xi}=-m\bm{\phi}
$ which is equivalent to 
\be
\varepsilon^{\alpha\beta}\pd_\alpha \xi_\beta=- m \Phi \,,
\ee
for the degree-zero component fields that we are interested in; this in turn gives  the Gauss law constraint albeit expressed via $E^\alpha$ rather than $D^\alpha$:
\be
\pd_\alpha E^\alpha=m B\,.
\ee
Alongside the usual position-momentum conjugate variables, we have therefore recovered the D2-brane phase space including the correct modified Gauss law arising from the topologically massive electrodynamics on the brane.

The preceding analysis is valid locally also in the case where all fluxes are switched on in \eqref{eq:D2Theta}. This is because whenever $Q^2=0$, the Hamiltonian with fluxes can be locally written
\be
\Theta= -\psi^\mu p_\mu +\tfrac{1}{2}m\phi^2 +\Theta_\text{flux}= -\psi^\mu p_\mu +\tfrac{1}{2}m\phi^2 + (-\psi^\mu p_\mu +\tfrac{1}{2}m\phi^2, \Psi) \,,
\ee
where $\Psi$ is the obvious function in degree $3$ including all relevant gauge potentials. The last equality states that the fluxes (excepting the Romans mass) can be removed by a local canonical transformation. The result for the brane phase space with fluxes included can thus be obtained from the fluxless result via a canonical transformation generated by the obvious expression involving potentials for all fluxes. This kind of argument is articulated in more detail in \cite[section 3.3]{Arvanitakis:2021wkt}.

\section{Discussion}\label{sec:discussion}

We introduced QP manifolds (also known or symplectic \lf-algebroids) that capture the exceptional generalised geometry of type IIA backgrounds including Romans mass \cite{Cassani:2016ncu,Ciceri:2016dmd}. Using canonical constructions that take QP manifolds as input, namely the AKSZ construction \cite{Alexandrov:1995kv} and the brane phase space construction \cite{Arvanitakis:2021wkt} we demonstrated how each of the ``D2'', ``D4'', and ``NS5'' algebroids corresponds to its namesake brane. This further establishes the connection between QP manifolds, generalised geometry, and string/M-theory. We also checked that the D2 and NS5 algebroids behave correctly with respect to type IIA string theory/M-theory duality: when the Romans mass vanishes, they arise from the M2 and M5 algebroids upon symplectic reduction with respect to a circle action, which corresponds to compactification of the 11th dimension on the string/M-theory side.

There remain a few puzzles with respect to the string theory interpretation. One is that all three algebroids can accommodate a closed 1-form flux $J_1$ (as in \eqref{eq:D2Theta} for the D2) on top of the expected NSNS and RR fluxes, which naturally transforms as a gauge connection for the $\mathbb R^+$ symmetry $\xi \rightarrow e^{\lambda}\xi$, $\phi \rightarrow e^{-\lambda} \phi$. This suggests a new IIA supergravity with a gauged $\bbR^{+}$ symmetry. While we can (classically) set $J_1=0$ by a canonical transformation, the $J_{1}$ field could define a genuinely different quantum theory. There are already known IIA theories with gauged $\bbR^{+}$ symmetry coming from the trombone symmetry \cite{Howe:1997qt,Lavrinenko:1997qa,Riccioni:2010xx}. However, the charges under the $\bbR^{+}$ symmetry of all the fields seem to not match and hence this is unlikely to correspond to the trombone symmetry. It would be interesting to investigate further whether one can define a consistent supergravity with these gauge symmetries.

Another puzzle is that the D2 and D4 algebroids accommodate Romans mass, but the NS5 algebroid does not. As we noted in the main text, this is consistent with the fact the D6-brane flux $F_8$ does not appear in the NS5 Q-structure because the mass and the D6 coupling appear together \cite{Bergshoeff:1997ak}. This ultimately underscores the continuing mystery of the absence of QP manifolds relevant for $(p>5)$-branes (equivalently, of degree $P>6$).

A perhaps-related mystery is the lack of known QP manifolds that encode information about the fermionic fields in supergravity. Generalised and exceptional generalised geometry and the closely related double/exceptional field theory seem to ``know'' information that would otherwise be obtained via supersymmetry; the relative coefficient between the kinetic and Wess-Zumino terms of the string is one example \cite{Arvanitakis:2017hwb,Arvanitakis:2018hfn}.  Our derivation of the D2-brane physics from (Romans massive) IIA supergravity data is another: it is a curious ``converse'' to the derivation of the Romans IIA supergravity equations from kappa-symmetry of the (topologically massive) D2-brane lagrangian \cite{Bergshoeff:1997cf}.

Finally, it would be interesting to relate the D4 algebroid to some of the other known ones via QP-manifold avatars of known stringy dualities. On the string theory side of course the D4 brane is obtained by e.g.~wrapping an M5 brane on the M-theory circle, or else via T-duality from the D3 brane. Fortunately a T-duality that relates Courant algebroids (i.e.~F1 algebroids in the terminology of this paper) has recently been realised \cite{Arvanitakis:2021lwo} in terms of (graded) Lagrangian correspondences. It therefore seems plausible there could exist a duality web linking the various QP-manifolds/generalised geometries which precisely mirrors the string/M-theory duality web.

\section*{Acknowledgements}
ASA acknowledges the support of the FWO-Vlaanderen through the project G006119N,  as well as that of the Vrije Universiteit Brussel through the Strategic Research Program ``High-Energy Physics''. He is also supported by an FWO Senior Postdoctoral Fellowship. EM is supported by the Deutsche Forschungsgemeinschaft (DFG, German Research Foundation) via the Emmy Noether program ``Exploring the landscape of string theory flux vacua using exceptional field theory'' (project number 426510644). DT is supported by the EPSRC New Horizons Grant ``New geometry from string dualities'' EP/V049089/1. We would like to thank Ond\v{r}ej Hul\'{i}k and Rahim Leung for helpful discussions, and Paul K.~Townsend for bringing interesting references to our attention.



\let\oldbibliography\thebibliography
\renewcommand{\thebibliography}[1]{%
	\oldbibliography{#1}%
	\setlength{\itemsep}{-1pt}%
}
\begin{multicols}{2}
{\setstretch{0}
	\small
	\bibliography{biblio.bib}}
\end{multicols}

\end{document}